\documentclass[%
 aip,
 jcp,%
 amsmath,amssymb,
 reprint,%
]{revtex4-1}

\usepackage{graphicx}
\usepackage{dcolumn}
\usepackage{bm}

\begin{document}

\title{Note: Variational Encoding of Protein Dynamics Benefits from Maximizing Latent Autocorrelation}

\author{Hannah K. Wayment-Steele}
 \affiliation{Department of Chemistry, Stanford University}
\author{Vijay S. Pande}
 \email{pande@stanford.edu}
\affiliation{Department of Bioengineering, Stanford University}

\date{\today}

\begin{abstract}
As deep Variational Auto-Encoder (VAE) frameworks become more widely used for modeling biomolecular simulation data, we emphasize the capability of the VAE architecture to concurrently maximize the timescale of the latent space while inferring a reduced coordinate, which assists in finding slow processes as according to the variational approach to conformational dynamics. We additionally provide evidence that the VDE framework (Hern\'{a}ndez et al., 2017), which uses this autocorrelation loss along with a time-lagged reconstruction loss, obtains a variationally optimized latent coordinate in comparison with related loss functions. We thus recommend leveraging the autocorrelation of the latent space while training neural network models of biomolecular simulation data to better represent slow processes.

\end{abstract}

\maketitle

\section*{}
The Variational Auto-Encoder (VAE) framework\cite{Kingma2013}, a neural network architecture for dimensionality reduction, is increasingly used for analyzing simulation data from biophysical systems\cite{hernandez2017variational, Wehmeyer2018, doerr2017dimensionality} and to infer collective variables for enhanced sampling simulations\cite{chen2017molecular, sultan2018transferable, ribeiro2018reweighted}. The VAE reduces high-dimensional data to a low-dimensional latent space by training two networks in parallel, one that ``encodes", or compresses the original data to a latent space, and one that ``decodes", or reconstructs the original data from the latent space. A loss function is used to train both networks concurrently. In the process of developing auto-encoder-based models for simulation data, several modifications to the original VAE loss function have been proposed to better suit the analysis of time-series data, and a more thorough analysis of the effect of these modifications on the modelled latent space is needed.

In this note, we compare the relative benefits of two modifications to the original VAE framework:~1) incorporating a loss term that encourages a latent coordinate with high autocorrelation, and 2) modifying the network to train on propagating data forward in time rather than reconstructing data at the same time point. We first describe how the standard VAE framework is applied to time-series data and then outline the two modifications above. We show that the first of these additions is essential for learning a meaningful latent coordinate for a real protein system, and that the combination of these two modifications, as first introduced in our recent work \cite{hernandez2017variational}, makes for a more highly autocorrelated latent coordinate, resulting in a better model according to the variational approach to conformational dynamics.

\hfill

\noindent{\bf The VAE applied to time-series data.} Consider a trajectory $x_t$ that we wish to compress, i.e.~encode, to a reduced-dimensionality latent space $z_t$. The traditional VAE framework learns a latent coordinate by iteratively 1) mapping the input coordinate $x_t$ to a latent space coordinate $z_t$ using the encoding network, $q_\phi(z_t | x_t )$, and 2) generating a reconstruction of the original coordinate, $\hat{x}_t$, using the decoding network, $p_\theta(\hat{x}_t | z_t) $. The standard VAE loss function (equation \ref{eq:VAE_loss}) trains both networks concurrently and is comprised of two terms. A reconstruction loss (which we will term $\mathcal{L}_\text{encod}$) aims to quantify how well the VAE reconstructs the data by minimizing the mean squared distance between the original data and the reconstructed data. A KL-divergence loss ($\mathcal{L}_\text{KL}$) on the encoded distribution, $q_\phi(z_t | x_t)$, and a prior on the latent encoding, $P(z)$, imposes a penalty on the complexity of the latent coordinate. This discourages the model from deterministically encoding each data point to a unique value, and instead encodes a distribution where neighboring points in the latent coordinate are encouraged to be correlated.


\begin{equation} \label{eq:VAE_loss}
\begin{split}
&\mathcal{L}_\text{VAE}(x_t; \theta, \phi) = \mathcal{L}_\text{encod} + \mathcal{L}_\text{KL} \\
&= \mathbb{E}_{p_\theta(\hat{x}_t | z_t)} \left[ (\hat{x}_t - x_t)^2 \right] + D_{KL} \left[ q_\phi (z_t | x_t) || P(z) \right].
\end{split}
\end{equation}

This standard framework can be augmented to better encode time-series data. We recently introduced the Variational Dynamics Encoder (VDE)\cite{hernandez2017variational}, which has been used both for analysis and in providing collective variables for enhanced sampling\cite{sultan2018transferable}. Our framework presented two modifications to the original VAE: 1) we incorporated a term to maximize the autocorrelation of the latent space, and 2) our decoder network was structured as a propagator, trained to reconstruct coordinates at some lag time in the future as opposed to reconstructing the input itself.

\hfill

\noindent{\bf A VAE can optimize the latent coordinate's timescale.}
Unlike other VAE-based methods for dimensionality reduction in biophysical simulation, our previous work\cite{hernandez2017variational} incorporates the autocorrelation of the latent coordinate in the loss function, which encourages the model to find a latent coordinate that is maximally autocorrelated,

\begin{equation}
\mathcal{L}_\rho =  - \rho_{z_t, z_{t+\tau}} = - \frac{\mathop{\mathbb{E}}\left[ \left(z_t - \bar{z}_t \right) \left( z_{t+\tau} - \bar{z}_{t+\tau} \right) \right]}{s_{z_t} s_{z_{t+\tau}}},
\end{equation}

where $\bar{z_t}$ is the batch mean of the encoded latent variable $z_t$ and $s_{z_t}$ is the batch standard deviation of $z_t$. This autocorrelation loss is motivated by the variational approach to conformational dynamics (VAC)\cite{noe2013}\footnote{NB: Though both use the variational principle, the VAC approach is distinct from the process of variational inference that the VAE framework performs, and for which the VAE is named}. The VAC states that in the limit of infinite sampling, no dynamical process can be approximated that is slower than the true slowest process. Thus, process timescales can be used can be used as an interpretation of model quality: by this reasoning, a model with slower processes is a better model of the system dynamics. Model quality may be evaluated using the maximizing the generalized matrix Raleigh quotient (GMRQ), a sum of the eigenvalues of a system given a decomposition to an approximated basis set. This has been employed in evaluating Markov State Models and has proven useful for parameter selection\cite{McGibbon2015,Husic2016}.

We can view the latent space of a VAE as a model that has the potential to identify the slowest process measurable through the expressivity afforded by neural networks. We can approximate the timescale of the latent space by measuring the autocorrelation $\rho$ of points in the latent space given a lag time $\tau$. Furthermore, the autocorrelation of the latent space is directly related to the sum of the eigenvalues, as shown in other works\cite{noe2011dynamical}. By including the autocorrelation of the latent space in our loss function, we are performing optimization on the quality of our model in representing long timescales, concurrently with using the VAE framework to perform variational inference to infer the latent coordinate $z_t$.

\begin{figure*}[t]
    \centering
    \includegraphics[width=0.8\textwidth]{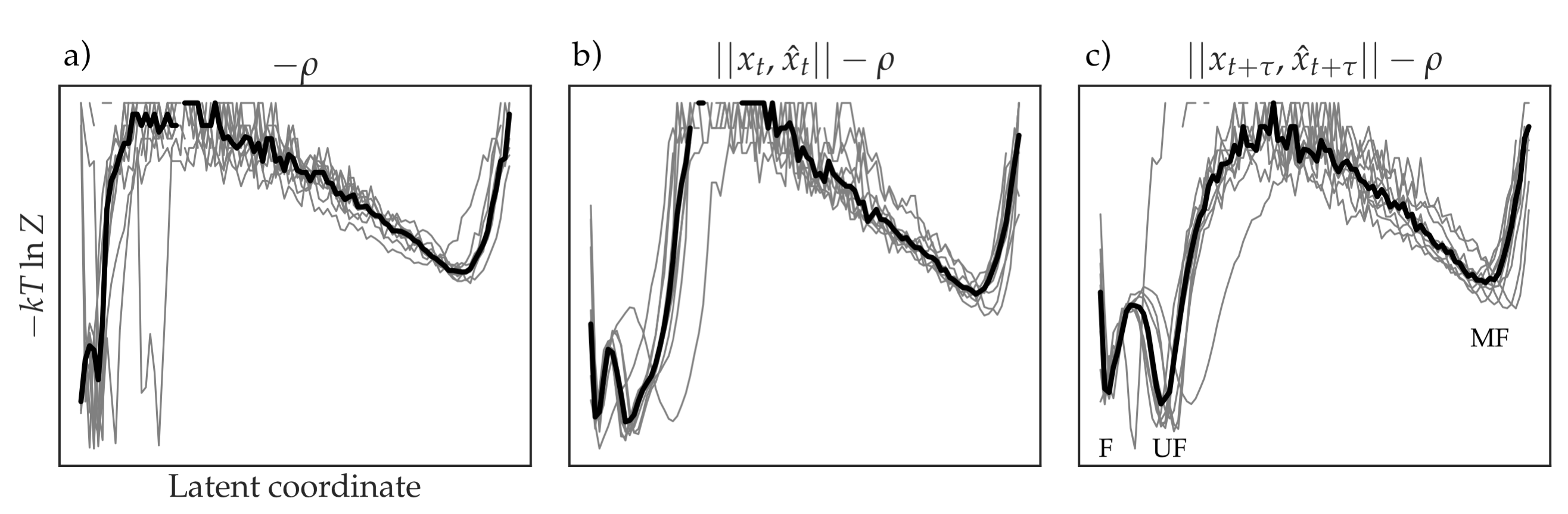}
    \caption{Free energies along latent coordinates identified by models trained with (a) autocorrelation loss $\mathcal{L}_\rho$, (b) encoding reconstruction loss and $\mathcal{L}_\rho$, (c) propagation reconstruction loss and $\mathcal{L}_\rho$.  Training on $\mathcal{L}_\rho$ alone (a) is insufficient to separate the folded and unfolded states (states labeled in c). The propagation reconstruction loss better separates these two states than the encoding reconstruction loss (labeled in c). Latent coordinates are averaged across 10 independently trained models for each condition. Individual model free energies are shown in grey and the average free energy per condition is shown in black.}
    \label{fig:free_energies}
\end{figure*}

\hfill

\noindent {\bf A VAE may be structured as a propagator.}
Our second modification to the original VAE framework structured the network as a propagator rather than an auto-encoder. This modification is used in other recently-developed VAE-based frameworks for time-series data as well\cite{Wehmeyer2018}. Instead of training a decoder to reconstruct the coordinate space given a datapoint at time $t$, we trained our decoder to reconstruct the coordinate space at time $t+\tau$, where $\tau$ is a user-selected lag time. This is mathematically analogous to the mean-squared error used in the standard VAE framework, denoted here as $\mathcal{L}_\text{encod}$. Formally, the VDE reconstruction loss is written as 
\begin{equation}
\mathcal{L}_\text{prop}(x_t; \theta) = \mathbb{E}_{p_\theta (\hat{x}_t | z_t)} \left[( \hat{x}_{t+\tau} - x_{t+\tau})^2\right].
\end{equation}
In this sense, our VDE network aims to approximate the propagator of the system, an operator that, given a distribution $f(x_t)$ is able to generate $f(x_{t+ \tau})$.


\hfill

\begin{figure*}
    \centering
    \includegraphics[width=\textwidth]{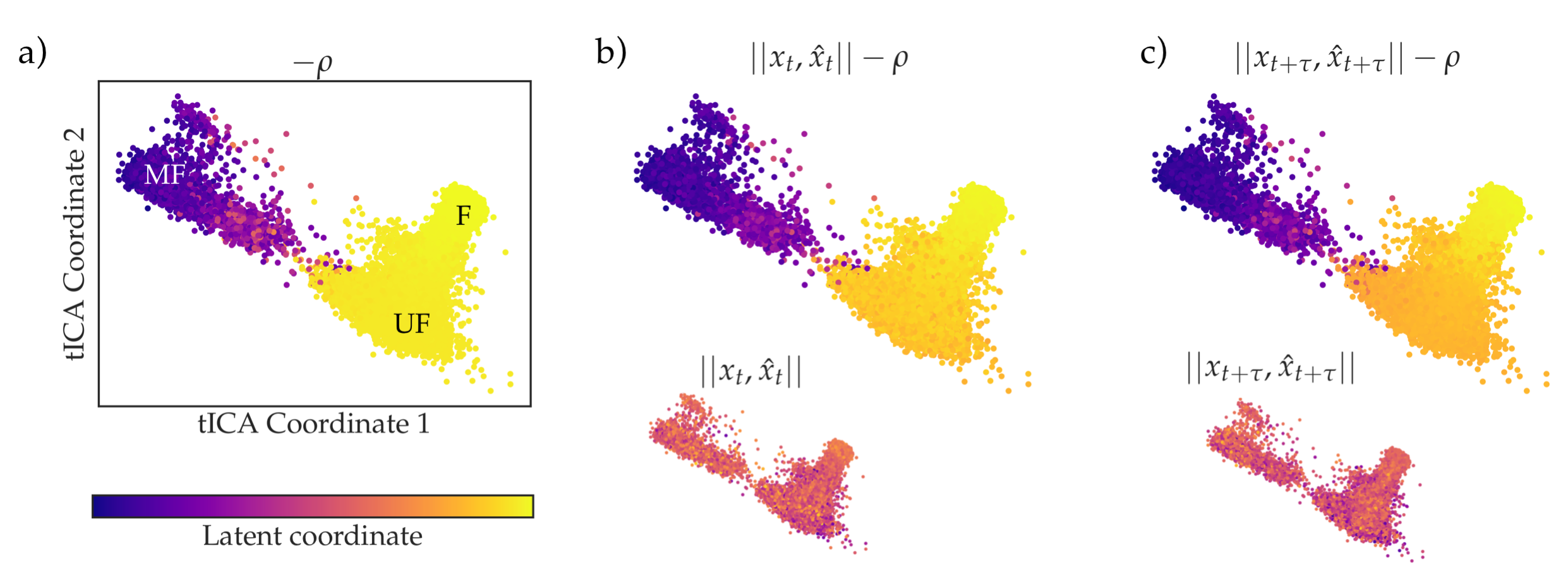}
    \caption{The latent variable autocorrelation loss ($\mathcal{L}_\rho$) is needed to obtain a useful encoding for the villin headpiece folding landscape. Here the landscape is plotted on the slowest two coordinates identified by linear tICA, and colored by VAE encodings with altered loss functions. Notable regions of the folding landscape include the folded (F), unfolded (UF), and a prominent misfolded state (MF), labeled in (a).  Training with $\mathcal{L}_\rho$ (a) is needed to encode the slow timescale of the landscape from MF to F/UF, and encoding reconstruction loss (b) or propagation reconstruction loss (c) further benefit the reconstruction, resolving the difference between the folded and unfolded state. Without the autocorrelation loss, neither the encoding reconstruction loss (b inset) nor the propagation reconstruction loss (c inset) is sufficient to encode the landscape meaningfully.}
    \label{fig:tica_projections}
\end{figure*}

\noindent {\bf Autocorrelation loss is needed to obtain a meaningful encoding.}
To analyze the effects of the modifications to the standard VAE loss function discussed above, the latent space autocorrelation loss $\mathcal{L}_\rho$ (equation 2) and propagation reconstruction loss ($\mathcal{L}_\text{prop}$ (equation 3), we compared models trained either with or without $\mathcal{L}_\rho$ and either with the standard VAE reconstruction loss, $\mathcal{L}_\text{encod}$, or $\mathcal{L}_\text{prop}$. As a control, we trained a model using only the encoding network trained only using $\mathcal{L}_\rho$, to isolate the effects of $\mathcal{L}_\rho$ from the two possible reconstruction losses $\mathcal{L}_\text{encod}$ and  $\mathcal{L}_\text{prop}$.

For each condition, we trained 10 independent models on simulation data\cite{lindorff2011fast} of the villin headpiece domain.  Each model was trained for 10 epochs using the parameters described previously\cite{hernandez2017variational}, with a lag time of 44 ns.  For all models that included $\mathcal{L}_\rho$, the training loss converged within 10 epochs and the models identified qualitatively very similar latent coordinates, while all models trained without $\mathcal{L}_\rho$ did not converge and did not find meaningful latent coordinates. In principle, VAE-based models trained without $\mathcal{L}_\rho$might converge to a similar latent coordinate given more training time, but we have not observed convergent behavior within 50 epochs and believe a framework that takes longer to train will not be useful in practice.

In figure 2, simulation data is projected onto the two slowest time-structured Independent Component Analysis (tICA) modes identified by an optimized tICA model \cite{Husic2016} for visualization purposes, and is colored by its projection onto the average latent coordinate identified for each loss function tested. We find that training with $\mathcal{L}_\rho$ alone is able to identify a latent coordinate separating the misfolded state (labeled MF in 1a) from the folded (F) and unfolded (UF) states. We again observe training with $\mathcal{L}_\rho$ alone is unable to separate the folded and unfolded state. Additionally incorporating either $\mathcal{L}_\text{encod}$ (1b) or $\mathcal{L}_\text{prop}$ (1c) result in a richer encoding that is able to separate the folded and unfolded state.  Importantly, we find that either a standard VAE loss function $\mathcal{L}_\text{encod}$ (1b, inset) or $\mathcal{L}_\text{prop}$ (1c, inset) are unable to find a meaningful latent coordinate.

To further probe the differences between latent coordinates trained using $\mathcal{L}_\rho$ alone and with $\mathcal{L}_\text{encod}$ or $\mathcal{L}_\text{prop}$, we plot the free energies of the resulting latent coordinates in figure 2. We observe that $\mathcal{L}_\rho$ alone does not differentiate the folded and unfolded state (2a).  Both $\mathcal{L}_\rho$ with $\mathcal{L}_\text{encod}$ or $\mathcal{L}_\text{prop}$ do, but the two states are more clearly differentiated when using $\mathcal{L}_\text{prop}$ (2c)  than with $\mathcal{L}_\text{encod}$ (2b).  

\hfill

\begin{figure}[ht]
    \centering
    \includegraphics[width=0.4\textwidth]{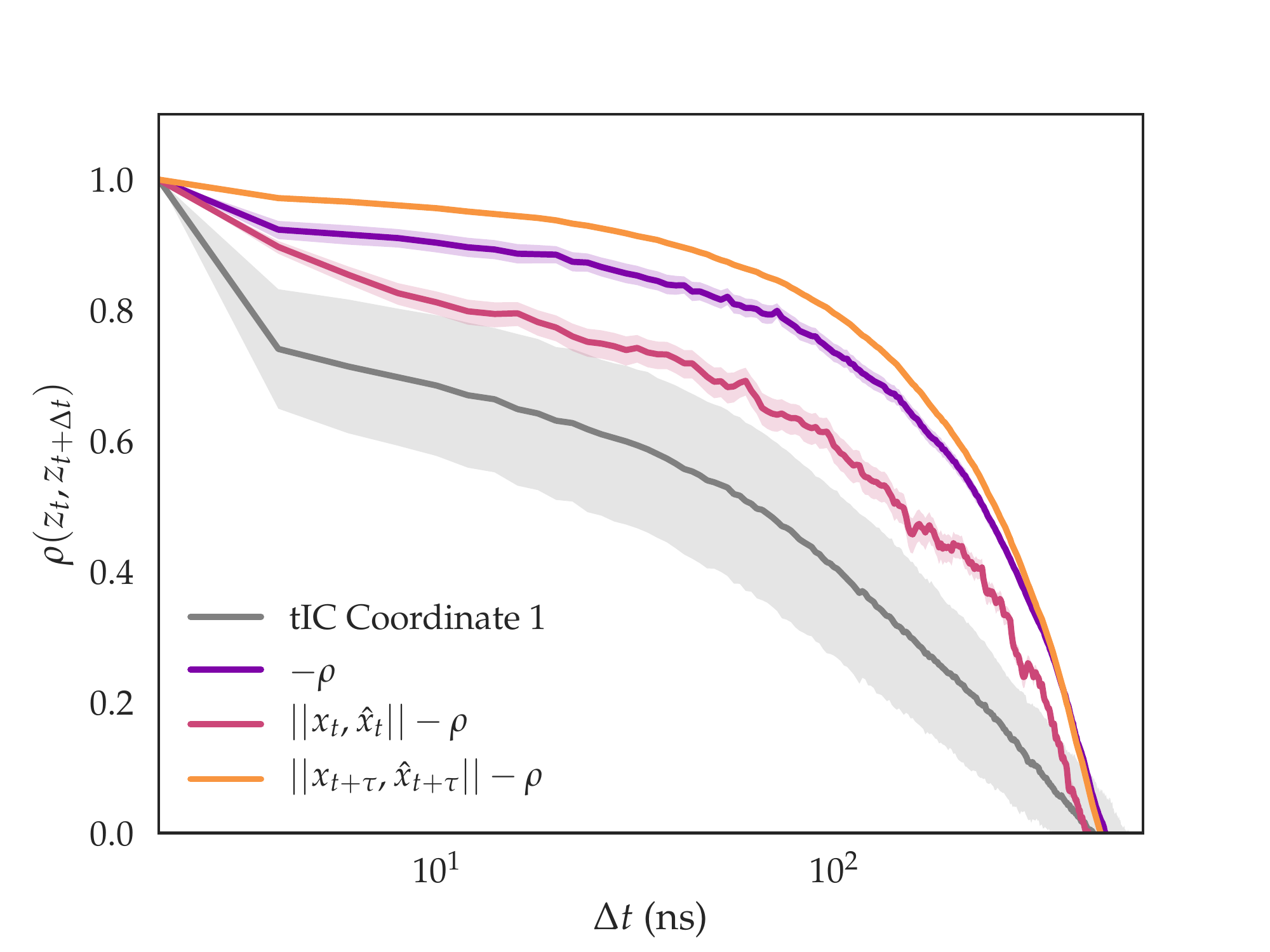}
    \caption{Autocorrelation of latent coordinates learned with varied loss functions. Using both  $\mathcal{L}_\rho$ and  $\mathcal{L}_\text{prop}$ (orange curve) results in a more autocorrelated latent coordinate than $\mathcal{L}_\rho$ alone (purple), whereas using $\mathcal{L}_\rho$  and the $\mathcal{L}_\text{encod}$ (pink) results in a less autocorrelated latent coordinate. For comparison, the autocorrelation of the first tICA coordinate is also shown (grey). Data is averaged over 10 independent models and all trajectories. Error bars represent the range of data observed.}
    \label{fig:autocorr}
\end{figure}

\noindent{\bf The VDE produces a variationally optimized model.}
To determine which loss function provided the most optimal model as defined by the variational principle for conformation dynamics, i.e.~the model identifying the process with the longest timescale, we computed the measured autocorrelations of each latent coordinate at a range of lag times.  The results are plotted in figure \ref{fig:autocorr}.  The model using $\mathcal{L}_\rho$ and $\mathcal{L}_\text{prop}$, the framework presented in the VDE\cite{hernandez2017variational}, has the most autocorrelated latent coordinate (orange curve), indicating a variationally optimal model out of the loss functions compared. Interestingly, the model using $\mathcal{L}_\rho$ and $\mathcal{L}_\text{prop}$ (pink curve) has lower autocorrelation than the model using only $\mathcal{L}_\rho$ (purple curve). We posit that the task of learning encoding ($\mathcal{L}_\text{encod}$) while jointly maximizing the autocorrelation of the latent space causes the model to overfit to the data at the expense of learning a highly autocorrelated latent variable, whereas the task of learning propagation ($\mathcal{L}_\text{prop}$), while more difficult, enables the network to learn a more autocorrelated latent coordinate.  Investigating this tradeoff is the topic of further study.

\hfill

In this work, we provide evidence that using the autocorrelation of the latent coordinate as a loss function ($\mathcal{L}_\rho$) is very useful and possibly essential for characterizing protein systems with VAE-based neural network models. Furthermore, we show that different types of reconstruction loss can either improve upon or detract from the quality of the model obtained with $\mathcal{L}_\rho$, as inferred by the autocorrelation of the latent space and the VAC. We observe that combining $\mathcal{L}_\rho$ with a standard VAE loss, which reconstructs $\hat{x}_t$ from $x_t$, results in a less-autocorrelated latent coordinate than using $\mathcal{L}_\rho$ only, whereas our modified reconstruction loss in the VDE which reconstructs $\hat{x}_{t+\tau}$  from $x_{t+\tau}$, results in a more-autocorrelated latent coordinate than $\mathcal{L}_\rho$ alone.

We note two technical points that prevent the guarantee that the measured implied timescale of the latent space is a true slowest process in the data. Firstly, the process of randomly selecting data batches for training any neural network-based model means that it is possible that for any given batch, the network is only trained to find a local timescale rather than the global longest timescale\cite{hernandez2017variational}. Secondly, to rigorously compare models by their dynamic process timescales, models should be cross-validated on the observed data to avoid overfitting to possible artifacts in the data\cite{McGibbon2015}.

As deep VAE models become more widely used for studying biophysical systems, we recommend including a loss term to maximize the autocorrelation of the latent space in variational auto-encoder frameworks. Doing so allows us to directly couple training the model with finding a latent representation with the longest possible timescale, leveraging existing theory regarding the variational principle for conformational dynamics and its implications for building optimal models.

\hfill

\noindent {\bf Acknowledgements}

The authors thank C. X. Hern\'{a}ndez, B. E. Husic, and M. M. Sultan for useful and insightful discussion. HKWS acknowledges support from NSF GRFP (DGE-114747). This work used the XStream computational resource, supported by the National Science Foundation Major Research Instrumentation program (ACI-1429830).

VSP is a consultant and SAB member of Schrodinger, LLC and Globavir, sits on the
Board of Directors of Apeel Inc, Freenome Inc, Omada Health, Patient Ping,
Rigetti Computing, and is a General Partner at Andreessen Horowitz.

\bibliography{paper}

\end{document}